\documentclass[a4paper,USenglish]{lipics-derivative}
\pdfoutput=1  % This tells arXiv to use pdflatex.
 
\hypersetup{hidelinks=true}

\usepackage{microtype}
\title{Asynchronous Distributed Automata: \protect\\
       A Characterization of the Modal Mu-Fragment}
\titlerunning{Asynchronous Distributed Automata}  % optional
\author[1]{Fabian Reiter}
\affil[1]{IRIF, Université Paris Diderot, France\\
  \texttt{fabian.reiter@gmail.com}}
\authorrunning{F. Reiter}
\Copyright{Fabian Reiter}  % http://creativecommons.org/licenses/by/3.0/
  % http://www.acm.org/about/class/ccs98-html
\EventEditors{}
\EventNoEds{0}
\EventLongTitle{}
\EventShortTitle{}
\EventAcronym{}
\EventYear{}
\EventDate{}
\EventLocation{}
\EventLogo{}
\SeriesVolume{}
\ArticleNo{}
\allowdisplaybreaks
\usepackage{mathtools}
\usepackage{stmaryrd}
\usepackage{xcolor}
\usepackage{xifthen}

\usepackage{bbm}

\usepackage{tikz}
\usetikzlibrary{automata,decorations.text}
\usepackage{standalone}

\usepackage[rightcaption]{sidecap}
\sidecaptionvpos{figure}{c}

\usepackage{sty/modallogic}
\usepackage{sty/commands}

\theoremstyle{plain}
\newtheorem{proposition}[theorem]{Proposition}

\begin{document}

\maketitle

\begin{abstract}
  We establish the equivalence
  between a class of asynchronous distributed automata
  and a small fragment of least fixpoint logic,
  when restricted to finite directed graphs.
  More specifically,
  the logic we consider is
  (a~variant~of) the fragment of the modal $\mu$-calculus
  that allows least fixpoints but forbids greatest fixpoints.
  The corresponding automaton model uses
  a network of identical finite-state machines
  that communicate in an asynchronous manner
  and whose state diagram must be acyclic except for self-loops.
  Exploiting the connection with logic,
  we also prove that the expressive power of those machines
  is independent of whether or not messages can be lost.
\end{abstract}

\section{Introduction}
One of the core disciplines of distributed computing
is to design and analyze message passing algorithms
that solve graph problems in computer networks.
Usually,
the problem instance considered in that context
is precisely the graph defined by the network
in which the computations are performed.
All nodes of the network run the same algorithm concurrently,
and often make no prior assumptions about the size and topology of the graph.
Typical problems that can be solved by such distributed algorithms include
graph coloring, leader election,
and the construction of spanning trees and maximal independent sets.
A comprehensive treatment of the subject can be found
in~\cite{DBLP:books/mk/Lynch96} and~\cite{Peleg00}.

The present paper follows up on
relatively recent results by Hella~et~al.\ and Kuusisto,
which establish novel connections between
modal logic and some restricted classes of distributed algorithms.
These weak types of algorithms,
referred to in the following as \emph{distributed automata},
can be represented as deterministic finite-state machines
that read sets of states instead of the usual alphabetic symbols.
Intuitively,
to run a distributed automaton on some node-labeled directed graph~$\Graph$,
a separate copy of the same machine is placed on every node
and initialized to a state that may depend on the node's label.
Each node~$\Node$ communicates with its peers
by sending its current state~$\State$ to every outgoing neighbor,
while at the same time collecting
the states received from its incoming neighbors
into a set~$\NeighborSet$.
The successor state of~$\State$ is then computed
as a function of~$\State$ and~$\NeighborSet$.
In particular,
this means that~$\Node$ cannot distinguish between
two incoming neighbors that share the same state.
Acting as a semi-decider,
the automaton accepts~$\Graph$ at position~$\Node$
precisely if~$\Node$ visits an accepting state at some point in time.
Either way,
all machines of the network run and communicate forever.

In~\cite{DBLP:conf/podc/HellaJKLLLSV12,DBLP:journals/dc/HellaJKLLLSV15},
Hella~et~al.\ have compared several classes of distributed algorithms,
of which the weakest uses the restricted communication model described above.
Deviating only in nonessential details from their original definition,
we can think of those weakest algorithms as
\emph{local synchronous} distributed automata.
Here,
“synchronous” means that all nodes of the network share a global clock,
thereby allowing the computation to proceed in an infinite sequence of rounds.
In each round,
all the nodes compute their next state simultaneously,
based on the information collected in the previous round.
By the term “local”
we mean that the nodes stop changing their state after a constant number of rounds,
a usage in accordance with the established terminology of distributed computing
(see, e.g., \cite{DBLP:journals/csur/Suomela13}).
Equivalently,
the state diagram of a local automaton is acyclic as long as we ignore sink states
(i.e., states that cannot be left once reached).
The work of Hella~et~al.\ reveals an intriguing link
between distributed computing and modal logic.
In particular,
it follows immediately
from~\mbox{\cite[Thm.~1]{DBLP:conf/podc/HellaJKLLLSV12,DBLP:journals/dc/HellaJKLLLSV15}}
that the graph properties recognizable by local synchronous automata
are precisely those definable in \emph{backward modal logic},
the variant of (basic) modal logic
where the usual modal operators are replaced by their backward-looking variants.

Motivated by the preceding result,
the connection with modal logic was further investigated by Kuusisto
in~\cite{DBLP:conf/csl/Kuusisto13} and~\cite{DBLP:journals/corr/Kuusisto14a}.
The former paper lifts the constraint of locality imposed
in~\cite{DBLP:conf/podc/HellaJKLLLSV12,DBLP:journals/dc/HellaJKLLLSV15},
thereby allowing automata with arbitrary state diagrams.
These (nonlocal) synchronous automata are then given a logical characterization
in terms of a new recursive logic dubbed \emph{modal substitution calculus}.
Furthermore,
\cite[Prp.~7]{DBLP:conf/csl/Kuusisto13} shows
that on finite graphs,
synchronous automata can easily recognize all the properties
definable in the least fixpoint fragment of the backward $\mu$-calculus.
This logic,
which we shall refer to simply as the \emph{backward $\mu$-fragment},
extends backward modal logic with a least fixpoint operator
that may not be negated.
It thus allows to express statements using least fixpoints,
but unlike in the full backward $\mu$-calculus,
greatest fixpoints are forbidden.
On the other hand,
the reverse conversion from synchronous automata to the backward $\mu$-fragment
is not possible in general.
As explained in \cite[Prp.~6]{DBLP:conf/csl/Kuusisto13},
it is easy to come up with a synchronous automaton
that makes crucial use of the fact
that a node can determine whether it receives the same information
from all of its incoming neighbors at exactly the same time.
Such a behavior cannot be simulated in the backward $\mu$-fragment.
By the same token,
even the much more expressive monadic second-order logic~(MSO)
is incomparable with synchronous automata.

Given that the preceding argument relies solely on synchrony,
it seems natural to ask
whether removing this feature can lead to a distributed automaton model
that has the same expressive power as the backward $\mu$-fragment.
The present paper provides a positive answer to this question.
We introduce several classes of \emph{asynchronous automata}
that transfer the standard notion of asynchronous algorithm
to the setting of finite-state machines.
Basically,
this means that we eliminate the global clock from the network,
thus making it possible for nodes to operate at different speeds
and for messages to be delayed for arbitrary amounts of time,
or even be lost.
From the syntactic point of view,
an asynchronous automaton is the same as a synchronous one,
but it has to satisfy an additional semantic condition:
its acceptance behavior must be independent of any timing-related issues.
Taking a closer look at the automata
obtained by translating formulas of the backward $\mu$-fragment,
we can easily see that they are in fact asynchronous.
Furthermore,
their state diagrams are almost acyclic,
except that all the states are allowed to have self-loops
(not only the sink states).
We call this property \emph{quasi-acyclic}.
The paper's main contribution is to show
that now we can also go in the other direction:
every quasi-acyclic asynchronous automaton
can be converted into an equivalent formula of the backward $\mu$-fragment.
Incidentally,
this remains true
even if we consider a seemingly more powerful variant of asynchronous automata,
where all messages are guaranteed to be delivered.
To illustrate the basic concepts,
an example of an automaton and an equivalent formula
will be provided in Figure~\ref{fig:automaton},
at the end of the next section.

The remainder of this paper is organized as follows:
After giving the necessary formal definitions in Section~\ref{sec:preliminaries},
we state and briefly discuss the main result in Section~\ref{sec:result}.
The proof is then developed in the last two sections.
Section~\ref{sec:logic-to-automata} presents
the rather straightforward translation from logic to automata.
The reverse translation is given in Section~\ref{sec:automata-to-logic},
which is a bit more involved
and therefore occupies the largest part of the paper.
\section{Preliminaries}
\label{sec:preliminaries}
We denote the set of Boolean values by $\defd{\Boolean} = \set{0,1}$,
the set of non-negative integers by $\defd{\Natural} = \set{0,1,2,\dots}$,
and the set of positive integers by $\defd{\Positive} = \Natural \setminus \set{0}$.
With respect to a given set $\Set$,
we write
\defd{$\powerset{\Set}$} for the power set,
\defd{$\Set^k$} for the set of $k$-tuples ($k \in \Natural$), and
\defd{$\card{\Set}$} for the cardinality.
As a special case of $k$-tuples,
$\Boolean^k$ denotes the set of all binary strings of length $k$.
Furthermore,
the length of a string $\String$ is written as \defd{$\length{\String}$}.

For $\BitCount \in \Natural$,
a (finite) \defd{$\BitCount$-bit labeled directed graph},
abbreviated \defd{\digraph{}},
is a structure $\Graph = \tuple{\NodeSet, \EdgeSet, \Labeling}$,
where
$\NodeSet$
is a finite nonempty set of nodes,
$\EdgeSet \subseteq \NodeSet \times \NodeSet$
is a set of directed edges, and
$\Labeling \colon \NodeSet \to \Boolean^\BitCount$
is a labeling that assigns a binary string of length $\BitCount$ to each node.
Isomorphic \digraph{s} are considered to be equal.
If $\Node$ lies in $\NodeSet$,
we call the pair~$\tuple{\Graph,\Node}$
a \defd{pointed \digraph{}}.
Moreover,
if $\Node[1]\Node[2]$ is an edge in $\EdgeSet$,
then $\Node[1]$ is called an \defd{incoming neighbor} of~$\Node[2]$.

\begin{definition}[Distributed Automaton]
  A (distributed) \defd{automaton}
  with $\BitCount$-bit input
  is a tuple $\Automaton = \tuple{\StateSet,\InitFunc,\TransFunc,\AcceptSet}$,
  where
  $\StateSet$
  is a finite set of states,
  $\InitFunc \colon \Boolean^\BitCount \to \StateSet$
  is an initialization function,
  $\TransFunc \colon \StateSet \times \powerset{\StateSet} \to \StateSet$
  is a transition function, and
  $\AcceptSet \subseteq \StateSet$
  is a set of accepting states.
\end{definition}

To run such an automaton~$\Automaton$ on a \digraph{}~$\Graph$,
we regard the edges of~$\Graph$ as FIFO buffers.
Each buffer~$\Node[2]\Node[3]$ will always contain a sequence of states
previously traversed by node~$\Node[2]$.
An adversary chooses
when $\Node[2]$ evaluates~$\TransFunc$
to push a new state to the back of the buffer,
and when the current first state gets popped from the front.
The details are clarified in the following.

A \defd{trace} of an automaton
$\Automaton = \tuple{\StateSet,\InitFunc,\TransFunc,\AcceptSet}$
is a finite nonempty sequence
$\Trace = \State_1 \dots \State_n$ of states in $\StateSet$
such that
$\State_i \neq \State_{i+1}$
and
$\TransFunc(\State_i,\NeighborSet_i) = \State_{i+1}$
for some $\NeighborSet_i \subseteq \StateSet$.
We say that $\Automaton$ is \defd{quasi-acyclic}
if its set of traces $\TraceSet$ is finite.
In other words, 
its state diagram must not contain any directed cycles,
except for self-loops.

For any states $\State[1],\State[2] \in \StateSet$ and
any (possibly empty) sequence $\Trace$ of states in $\StateSet$,
we define the unary postfix operators
\defd{$\first$}, \defd{$\last$}, \defd{$\pushlast$} and \defd{$\popfirst$}
as follows:
$\State[1]\Trace.\first = \Trace\State[1].\last = \State[1]$,
\begin{equation*}
  \Trace\State[1].\pushlast(\State[2]) =
  \begin{cases*}
    \Trace\State[1]\State[2] & if $\State[1] \neq \State[2]$, \\
    \Trace\State[1]          & if $\State[1] = \State[2]$,
  \end{cases*}
  \qquad
  \text{and}
  \qquad
  \State[1]\Trace.\popfirst =
  \begin{cases*}
    \Trace          & if $\Trace$ is nonempty, \\
    \State[1]\Trace & if $\Trace$ is empty.
  \end{cases*}
\end{equation*}

An (asynchronous) \defd{timing} of a \digraph{}
$\Graph = \tuple{\NodeSet, \EdgeSet, \Labeling}$
is an infinite sequence
$\Timing = \tuple{\Timing_1, \Timing_2, \Timing_3, \dots}$
of maps
$\Timing_\Time \colon \NodeSet \cup \EdgeSet \to \Boolean$,
indicating which nodes and edges are active at time~$\Time$,
where $1$ is assigned infinitely often
to every node and every edge.
More formally,
for all $\Time[1] \in \Positive$, $\Node \in \NodeSet$ and $\Edge \in \EdgeSet$,
there exist $\Time[2],\Time[3] > \Time[1]$
such that $\Timing_{\Time[2]}(\Node) = 1$ and $\Timing_{\Time[3]}(\Edge) = 1$.
We refer to this as the \defd{fairness property} of $\Timing$.
As a restriction,
we say that $\Timing$ is \defd{lossless-asynchronous}
if $\Timing_\Time(\Node[1]\Node[2]) = 1$ implies $\Timing_\Time(\Node[2]) = 1$
for all $\Time \in \Positive$ and $\Node[1]\Node[2] \in \EdgeSet$.
Furthermore,
$\Timing$ is called the (unique) \defd{synchronous timing} of $\Graph$
if $\Timing_\Time(\Node) = \Timing_\Time(\Edge) = 1$
for all $\Time \in \Positive$, $\Node \in \NodeSet$ and $\Edge \in \EdgeSet$.

\begin{definition}[Asynchronous Run]
  Let
  $\Automaton = \tuple{\StateSet,\InitFunc,\TransFunc,\AcceptSet}$
  be a distributed automaton with \mbox{$\BitCount$-bit} input
  and $\TraceSet$ be its set of traces.
  Furthermore, let
  $\Graph = \tuple{\NodeSet, \EdgeSet, \Labeling}$
  be an $\BitCount$-bit labeled \digraph{}
  and
  $\Timing = \tuple{\Timing_1, \Timing_2, \Timing_3, \dots}$
  be a timing of $\Graph$.
  The (asynchronous) \defd{run} of $\Automaton$ on~$\Graph$ timed by $\Timing$
  is the infinite sequence
  $\Run = \tuple{\Run_0, \Run_1, \Run_2, \dots}$
  of \defd{configurations}
  $\Run_\Time \colon \NodeSet \cup \EdgeSet \to \TraceSet$,
  with $\Run_\Time(\NodeSet) \subseteq \StateSet$,
  which are defined inductively as follows,
  for $\Time \in \Natural$, $\Node[2] \in \NodeSet$ and $\Node[2]\Node[3] \in \EdgeSet$:
  \begin{align*}
    \Run_0(\Node[2]) &= \Run_0(\Node[2]\Node[3]) = \InitFunc(\Labeling(\Node[2])), \\[1ex]
    \Run_{\Time+1}(\Node[2]) &=
    \begin{cases*}
      \swl{\Run_\Time(\Node[2])}{\Run_\Time(\Node[2]\Node[3]).\pushlast(\Run_{\Time+1}(\Node[2])).\popfirst}
        & if $\Timing_{\Time+1}(\Node[2]) = 0$, \\
      \TransFunc \bigl( \Run_\Time(\Node[2]),
                        \setbuilder{\Run_\Time(\Node[1]\Node[2]).\first}{\Node[1]\Node[2] \in \EdgeSet}
                 \bigr)
        & if $\Timing_{\Time+1}(\Node[2]) = 1$,
    \end{cases*} \\[1ex]
    \Run_{\Time+1}(\Node[2]\Node[3]) &=
    \begin{cases*}
      \Run_\Time(\Node[2]\Node[3]).\pushlast(\Run_{\Time+1}(\Node[2]))
        & if $\Timing_{\Time+1}(\Node[2]\Node[3]) = 0$, \\
      \Run_\Time(\Node[2]\Node[3]).\pushlast(\Run_{\Time+1}(\Node[2])).\popfirst
        & if $\Timing_{\Time+1}(\Node[2]\Node[3]) = 1$.
    \end{cases*}
  \end{align*}
  If $\Timing$ is the synchronous timing of $\Graph$,
  we refer to $\Run$ as the \defd{synchronous run} of $\Automaton$ on $\Graph$.
\end{definition}

Throughout this paper,
we assume that our \digraph{s}, automata and logical formulas
agree on the number $\BitCount$ of labeling bits.
An automaton $\Automaton$ \defd{accepts}
a pointed \digraph{} $\tuple{\Graph,\Node}$
under timing $\Timing$
if $\Node$ visits an accepting state at some point
in the run $\Run$ of $\Automaton$ on $\Graph$ timed by $\Timing$,
i.e., if there exists $\Time \in \Natural$
such that $\Run_\Time(\Node) \in \AcceptSet$.
If we simply say that $\Automaton$ accepts $\tuple{\Graph,\Node}$,
without explicitly specifying a timing $\Timing$,
then we stipulate that $\Run$ is the synchronous run of $\Automaton$ on $\Graph$.

Given a \digraph{}
$\Graph = \tuple{\NodeSet, \EdgeSet, \Labeling}$
and a class $\TimingSet$ of timings of $\Graph$,
the automaton $\Automaton$ is called \defd{consistent}
for $\Graph$ and $\TimingSet$
if for all $\Node \in \NodeSet$,
either $\Automaton$ accepts $\tuple{\Graph,\Node}$
under every timing in $\TimingSet$,
or $\Automaton$ does not accept $\tuple{\Graph,\Node}$
under any timing in $\TimingSet$.
We say that $\Automaton$ is \defd{asynchronous}
if it is consistent for every possible choice of $\Graph$ and $\TimingSet$,
and \defd{lossless-asynchronous}
if it is consistent for every choice
where $\TimingSet$ contains only lossless-asynchronous timings.
By contrast,
we call an automaton \defd{synchronous}
if we wish to emphasize
that no such consistency requirements are imposed.
Intuitively,
all automata can operate in the synchronous setting,
but only some of them also work reliably
in environments that provide fewer guarantees.

A \defd{\digraph{} property} is a set $\Property$ of pointed \digraph{s}.
We call $\Property$ the \digraph{} property
\defd{recognized} by an automaton $\Automaton$
if it consist precisely of those pointed \digraph{s}
that are accepted by $\Automaton$.
We denote by \defd{$\AA$}, \defd{$\LA$} and \defd{$\SA$}
the classes of \digraph{} properties recognizable by 
asynchronous, lossless-asynchronous and synchronous automata, respectively.
Similarly,
\defd{$\QAA$}, \defd{$\QLA$} and \defd{$\QSA$}
are the corresponding classes recognizable by quasi-acyclic automata.

Turning to logic,
let \defd{$\VariableSet$} be an infinite supply
of propositional variables.
We define the formulas of \defd{backward modal logic}
with~$\BitCount$ propositional constants
by means of the grammar
\begin{equation*}
  \Formula \Coloneqq \bot
                \mid \top
                \mid \Constant_i
                \mid \lnot \Constant_i
                \mid \Variable
                \mid (\Formula \lor \Formula)
                \mid (\Formula \land \Formula)
                \mid \bdm \Formula
                \mid \bbx \Formula \,,
\end{equation*}
where $0 \leq i < \BitCount$ and $\Variable \in \VariableSet$.
Note that this syntax ensures that variables cannot be negated.
Given such a formula $\Formula$,
an $\BitCount$-bit labeled \digraph{}
$\Graph = \tuple{\NodeSet, \EdgeSet, \Labeling}$
and a variable assignment
$\Assignment \colon \VariableSet \to \powerset{\NodeSet}$,
we write
\defd{$\sem[\Graph,\Assignment]{\Formula}$}
to denote the subset of nodes of~$\Graph$
at which $\Formula$ holds
with respect to $\Assignment$.
For \defd{atomic propositions} $\Constant_i$ and $\Variable$,
the corresponding semantics are defined by
$\sem[\Graph,\Assignment]{\Constant_i}
= \setbuilder{\Node \in \NodeSet}{\Labeling(\Node)(i) = 1}$
and
$\sem[\Graph,\Assignment]{\Variable} = \Assignment(\Variable)$,
where $\Labeling(\Node)(i)$ is the $i$-th bit of $\Labeling(\Node)$.
The Boolean constants and connectives are interpreted in the usual way,
for instance,
$\sem[\Graph,\Assignment]{\top} = \NodeSet$
and
$\sem[\Graph,\Assignment]{(\Formula[1] \lor \Formula[2])} =
\sem[\Graph,\Assignment]{\Formula[1]} \cup \sem[\Graph,\Assignment]{\Formula[2]}$.
Finally,
the \defd{backward diamond}~$\bdm$ and the \defd{backward box}~$\bbx$
represent backward-looking modal operators,
with the semantics
\begin{align*}
  \sem[\Graph,\Assignment]{\bdm \Formula} &=
  \lrsetbuilder{\Node[2] \in \NodeSet}
               {\text{$\Node[1] \in \sem[\Graph,\Assignment]{\Formula}$
                      for some $\Node[1] \in \NodeSet$
                      such that $\Node[1]\Node[2] \in \EdgeSet$}}
  \quad \text{and} \\
  \sem[\Graph,\Assignment]{\bbx \Formula} &=
  \lrsetbuilder{\Node[2] \in \NodeSet}
               {\text{$\Node[1] \in \sem[\Graph,\Assignment]{\Formula}$
                      for all $\Node[1] \in \NodeSet$
                      such that $\Node[1]\Node[2] \in \EdgeSet$}}.
\end{align*}

Traditionally,
the modal $\mu$-calculus is defined to comprise individual fixpoints
which may be nested.
However,
it is well-known that we can add simultaneous fixpoints
to the $\mu$-calculus
without changing its expressive power,
and that nested fixpoints of the same type (i.e., least or greatest)
can be rewritten as non-nested simultaneous ones
(see, e.g., \cite[\S~3.7]{BradfieldS07} or \cite[\S~4.3]{Lenzi05}).
The following definition directly takes advantage of this fact.
We shall restrict ourselves to the
\defd{$\mu$-fragment of the backward $\mu$-calculus},
abbreviated \defd{backward $\mu$-fragment},
where only least fixpoints are allowed,
and where the usual modal operators are replaced by their backward-looking variants.
Without loss of generality,
we stipulate that each formula of
the backward $\mu$-fragment with $\BitCount$ propositional constants
is of the form
\begin{equation*}
  \Formula \:=\: \mu \!
  \begin{pmatrix}
    \Variable_0 \\
    \vdots \\
    \Variable_\VarMax
  \end{pmatrix}
  \! . \!
  \begin{pmatrix}
    \Formula_0(\Constant_0, \dots, \Constant_{\BitCount-1}, \Variable_0, \dots, \Variable_\VarMax) \\
    \vdots \\
    \Formula_\VarMax(\Constant_0, \dots, \Constant_{\BitCount-1}, \Variable_0, \dots, \Variable_\VarMax)
  \end{pmatrix},
\end{equation*}
where
$\Variable_0,\dots,\Variable_\VarMax \in \VariableSet$,
and $\Formula_0,\dots,\Formula_\VarMax$ are formulas of
backward modal logic with $\BitCount$ propositional constants
that may contain no other variables than $\Variable_0, \dots, \Variable_\VarMax$.

For every \digraph{}
$\Graph = \tuple{\NodeSet, \EdgeSet, \Labeling}$,
the tuple $\tuple{\Formula_0,\dots,\Formula_\VarMax}$
gives rise to an operator
$\Operator \colon (\powerset{\NodeSet})^{\VarMax+1} \to (\powerset{\NodeSet})^{\VarMax+1}$
that takes some valuation of
$\vec{\Variable} = \tuple{\Variable_0,\dots,\Variable_\VarMax}$
and reassigns to each $\Variable_i$ the resulting valuation of $\Formula_i$.
More formally,
$\Operator$~maps $\vec{\NodeSubset} = \tuple{\NodeSubset_0,\dots,\NodeSubset_\VarMax}$
to $\tuple{\NodeSubset'_0,\dots,\NodeSubset'_\VarMax}$
such that
$\NodeSubset'_i =
 \sem[\Graph,\ver{}{\vec{\Variable}}{\vec{\NodeSubset}}]{\Formula_i}$.
Here,
$\ver{}{\vec{\Variable}}{\vec{\NodeSubset}}$
can be any variable assignment
that interprets each $\Variable_i$~as~$\NodeSubset_i$.
A~(simultaneous) \defd{fixpoint} of the operator~$\Operator$
is a tuple
$\vec{\NodeSubset} \in (\powerset{\NodeSet})^{\VarMax+1}$
such that
$\Operator(\vec{\NodeSubset}) = \vec{\NodeSubset}$.
Since, by definition,
variables occur only positively in formulas,
the operator $\Operator$ is \defd{monotonic}.
This means that
$\vec{\NodeSubset} \subseteq \vec{\NodeSubset}'$
implies
$\Operator(\vec{\NodeSubset}) \subseteq \Operator(\vec{\NodeSubset}')$
for all $\vec{\NodeSubset},\vec{\NodeSubset}' \in (\powerset{\NodeSet})^{\VarMax+1}$,
where set inclusions are to be understood componentwise
(i.e., $\NodeSubset_i \subseteq \NodeSubset'_i$ for each $i$).
Therefore,
by virtue of a theorem due to Knaster and Tarski,
$\Operator$~has a \mbox{\defd{least fixpoint}},
which is defined as the unique fixpoint
$\vec{\LFixpoint} = \tuple{\LFixpoint_0,\dots,\LFixpoint_\VarMax}$
of~$\Operator$
such that
$\vec{\LFixpoint} \subseteq \vec{\NodeSubset}$
for every other fixpoint $\vec{\NodeSubset}$ of $\Operator$.
As a matter of fact,
the Knaster-Tarski theorem even tells us that
$\vec{\LFixpoint}$ is equal to
$\bigcap \setbuilder{\vec{\NodeSubset} \in (\powerset{\NodeSet})^{\VarMax+1}}
                    {\Operator(\vec{\NodeSubset}) \subseteq \vec{\NodeSubset}}$,
where set operations must also be understood componentwise.
Another, perhaps more intuitive, way of characterizing $\vec{\LFixpoint}$
is to consider the inductively constructed sequence of approximants
$\tuple{\vec{\LFixpoint}^0, \vec{\LFixpoint}^1, \vec{\LFixpoint}^2, \dots}$,
where
$\vec{\LFixpoint}^0 = \tuple{\emptyset, \dots, \emptyset}$
and
$\vec{\LFixpoint}^{j+1} = \Operator(\vec{\LFixpoint}^j)$.
Since this sequence is monotonically increasing and $\NodeSet$ is finite,
there exists $n \in \Natural$ such that
$\vec{\LFixpoint}^n = \vec{\LFixpoint}^{n+1}$.
It is easy to check that
$\vec{\LFixpoint}^n$ coincides with the least fixpoint $\vec{\LFixpoint}$.
For more details and proofs, see, e.g.,
\cite[\S~3.3.1]{DBLP:series/txtcs/GradelKLMSVVW07}.

\begin{SCfigure}
  \begin{tikzpicture}[->, semithick, auto, node distance=25ex]
  \tikzset{every state/.append style={inner sep=0ex, minimum size=6.9ex}}
  \tikzset{every edge/.append style={inner sep=0.5ex,font=\footnotesize}}
  \tikzset{every loop/.style={inner sep=0.3ex,looseness=4}}
  \node[initial,state,initial text={\normalsize $1$},initial distance=2ex,align=center]
    (1)
    {$1$ \\[-1ex] $\scriptstyle (\Constant_0)$};
  \node[initial,state,initial text={\normalsize $0$},initial distance=2ex]
    (2) [below of=1,yshift=3ex]
    {$2$};
  \node[accepting,state,align=center]
    (3) [right of=1]
    {$3$ \\[-1ex] $\scriptstyle (\Variable[1])$};
  \node[state,align=center]
    (4) [right of=2]
    {$4$ \\[-1ex] $\scriptstyle (\Variable[2])$};
  \node[accepting,state,align=center]
    (5) [right of=3,xshift=-2.5ex]
    {$5$ \\[-1ex] $\scriptstyle (\Variable[1],\Variable[2])$};

  \path (1) edge [loop below] node {otherwise} (1)
            edge              node [align=center]
                                   {if $\NeighborSet \nsubseteq \set{4,5}$ \\
                                    and $\NeighborSet \nsubseteq \set{1,2,4}$} (3)
        (2) edge [loop below] node {otherwise} (2)
            edge              node [sloped,anchor=south,pos=0.38,align=left]
                                   {if $\NeighborSet \nsubseteq \set{4,5}$ \\
                                    and $\NeighborSet \nsubseteq \set{1,2,4}$} (3)
            edge              node [swap,pos=0.42]
                                   {if $\NeighborSet \subseteq \set{4}$} (4)
            edge              node [sloped,swap,anchor=north,pos=0.28]
                                   {if $\set{5} \subseteq \NeighborSet \subseteq \set{4,5}$} (5)
        (3) edge [loop below] node {otherwise} (3)
            edge              node [align=left]
                                   {if $\NeighborSet \subseteq \set{4,5}$} (5)
        (4) edge [loop below] node {otherwise} (4)
            edge              node [sloped,anchor=north,pos=0.29]
                                   {if $\set{5} \subseteq \NeighborSet$} (5)
        (5) edge [loop below] node {always} (5);
  \draw [postaction={decorate,decoration={raise=1ex,text along path,text align={left,left indent=2.6ex},
                                          text={|\footnotesize|if {$\NeighborSet \subseteq \set{4,5}$}}}}]
        (1) to [bend left=39] (5);
  \node [right of=4,xshift=-6ex,yshift=0ex,align=right,font={\small\itshape}]
        {$\NeighborSet$: set of \\ received \\ states};
\end{tikzpicture}
  \caption{
    A quasi-acyclic asynchronous distributed automaton
    equivalent to the formula
    $$\mu \!
    \begin{pmatrix}
      \Variable[1] \\[0.5ex]
      \Variable[2]
    \end{pmatrix}
    \! . \!
    \begin{pmatrix}
      (\Constant_0 \land \Variable[2]) \,\lor\, \bdm \Variable[1] \\[0.5ex]
      \bbx \, \Variable[2]
    \end{pmatrix}$$
    of the backward $\mu$-fragment.
    A given pointed $1$-bit labeled \digraph{} $\tuple{\Graph,\Node[2]}$
    is accepted by this automaton
    if and only if,
    starting at~$\Node[2]$
    and following $\Graph$'s edges in the backward direction,
    it is possible to reach some node~$\Node[1]$ labeled with~$1$
    from which it is impossible to reach any directed cycle.
  }
  \label{fig:automaton}
\end{SCfigure}

Having introduced the necessary background,
we can finally establish the semantics of~$\Formula$
with respect to $\Graph$:
the set~\defd{$\sem[\Graph]{\Formula}$} of nodes at which $\Formula$ holds
is precisely~$\LFixpoint_0$,
the first component of~$\vec{\LFixpoint}$.
A pointed \digraph{}~$\tuple{\Graph,\Node}$ satisfies~$\Formula$,
in symbols \defd{$\tuple{\Graph,\Node} \models \Formula$},
if $\Node \in \sem[\Graph]{\Formula}$.
Accordingly,
the \digraph{} property \defd{defined} by~$\Formula$ is
$\setbuilder{\tuple{\Graph,\Node}}{\tuple{\Graph,\Node} \models \Formula}$,
and we denote by~\defd{$\SigmaOne$} the class of all \digraph{} properties
defined by some formula of the backward $\mu$-fragment.

As usual,
two devices (i.e., automata or formulas) are \defd{equivalent}
if they specify (i.e., recognize or define) the same property.
Figure~\ref{fig:automaton} provides an example of such an equivalence.
\section{Main result}
\label{sec:result}
\begin{figure}
  \centering
  \begin{subfigure}[t]{0.35\textwidth}
    \centering
    \begin{tikzpicture}[->, semithick, node distance=7ex]
  \node (SA) {$\SA$};
  \node (LA) [below of=SA] {$\LA$};
  \node (AA) [below of=LA] {$\AA$};
  \node (QSA) [right of=SA,xshift=10ex] {$\QSA$};
  \node (QLA) [below of=QSA] {$\QLA$};
  \node (QAA) [below of=QLA] {$\QAA$};
  \path
    (QSA) edge (SA)
    (QLA) edge (LA)
    (QAA) edge (AA)
    (AA)  edge (LA)
    (LA)  edge (SA)
    (QAA) edge (QLA)
    (QLA) edge (QSA);
\end{tikzpicture}
    \caption{immediate by the definitions}
    \label{fig:before-result}
  \end{subfigure}
  \qquad
  \begin{subfigure}[t]{0.4\textwidth}
    \centering
    \begin{tikzpicture}[->, semithick, node distance=7ex]
  \node (SA) {$\SA$};
  \node (LA) [below of=SA] {$\LA$};
  \node (AA) [below of=LA] {$\AA$};
  \node (QSA) [right of=SA,xshift=10ex] {$\QSA$};
  \node (QLA) [below of=QSA] {};
  \node (QAA) [below of=QLA] {$\SigmaOne = \QAA = \QLA$};
  \path
    (QSA) edge (SA)
    (QAA) edge (AA)
    (AA)  edge (LA)
    (LA)  edge (SA)
    (QAA) edge (QSA);
\end{tikzpicture}
    \caption{collapse shown in this paper}
    \label{fig:after-result}
  \end{subfigure}
  \caption{
    Hierarchy of the classes of \digraph{} properties
    recognizable by distributed automata,
    depending on whether the automata are
    synchronous~(S),
    lossless-asynchronous~(L),
    asynchronous~(A), or
    quasi-acyclic~(Q).
    The arrows denote inclusion
    (e.g., $\LA \subseteq \SA$).
  }
  \label{fig:result}
\end{figure}
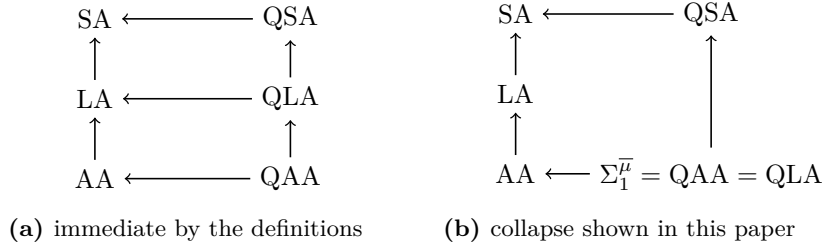

Based on the definitions given in Section~\ref{sec:preliminaries},
asynchronous automata are a special case of lossless-asynchronous automata,
which in turn are a special case of synchronous automata.\footnote{
  This may seem counterintuitive at first sight,
  but it is actually consistent
  with the standard terminology of distributed computing:
  an asynchronous algorithm can always serve as a synchronous algorithm
  (i.e., it can be executed in a synchronous environment),
  but the converse is not true.
}
Furthermore,
quasi-acyclicity constitutes an additional
(possibly orthogonal) restriction on these models.
We thus immediately obtain the hierarchy of classes
depicted in Figure~\ref{fig:before-result}.

Our main result provides a simplification of this hierarchy:
the classes $\QAA$ and $\QLA$ are actually equal to
the class of \digraph{} properties definable in the backward $\mu$-fragment.
This yields the revised diagram shown in Figure~\ref{fig:after-result}.

\begin{theorem}[$\SigmaOne = \QAA = \QLA$]
  \label{thm:main-result}
  When restricted to finite \digraph{s},
  the backward \mbox{$\mu$-fragment}
  is effectively equivalent to the classes of
  quasi-acyclic asynchronous automata and
  quasi-acyclic lossless-asynchronous automata.
\end{theorem}
\begin{proof}
  The forward direction is given by
  Proposition~\ref{prp:logic-to-automata} (in Section~\ref{sec:logic-to-automata}),
  which asserts that $\SigmaOne \subseteq \QAA$,
  and the trivial observation that $\QAA \subseteq \QLA$.
  For the backward direction,
  we use Proposition~\ref{prp:automata-to-logic} (in Section~\ref{sec:automata-to-logic}),
  which asserts that $\QLA \subseteq \SigmaOne$.
\end{proof}

As stated before,
synchronous automata are more powerful than the backward $\mu$-fragment
(and incomparable with monadic second-order logic).
This holds even if we consider only quasi-acyclic automata,
i.e., the inclusion $\SigmaOne \subset \QSA$ is known to be strict
(see \mbox{\cite[Prp.~6]{DBLP:conf/csl/Kuusisto13}}).
Moreover,
an upcoming paper will show that
the inclusion $\QSA \subset \SA$ is also strict.

In contrast,
it remains open whether quasi-acyclicity
is in fact necessary for characterizing $\SigmaOne$.
On the one hand,
this notion is crucial for our proof
(see Proposition~\ref{prp:automata-to-logic}),
but on the other hand,
no \digraph{} property separating $\AA$ or $\LA$ from $\SigmaOne$
has been found so far.
\section{Computing least fixpoints using asynchronous automata}
\label{sec:logic-to-automata}
In this section,
we prove the easy direction of the main result.
Given a formula~$\Formula$ of the backward $\mu$-fragment,
it is straightforward to construct
a (synchronous) distributed automaton~$\Automaton$
that computes on any \digraph{} the least fixpoint~$\vec{\LFixpoint}$
of the operator associated with~$\Formula$.
As long as it operates in the synchronous setting,
$\Automaton$ simply follows the sequence of approximants
$\tuple{\vec{\LFixpoint}^0, \vec{\LFixpoint}^1, \dots}$
described in Section~\ref{sec:preliminaries}.
It is important to stress that
the very same observation has previously been made
in~\cite[Prp.~7]{DBLP:conf/csl/Kuusisto13}
(formulated from a different point of view).
In the following proposition,
we refine this observation
by giving a more precise characterization of the obtained automaton:
it is always quasi-acyclic and
capable of operating in a (possibly lossy) asynchronous environment.

\begin{proposition}[$\SigmaOne \subseteq \QAA$]
  \label{prp:logic-to-automata}
  For every formula of the backward $\mu$-fragment,
  we can effectively construct
  an equivalent quasi-acyclic asynchronous automaton.
\end{proposition}
\begin{proof}
  Let 
  $\Formula = \mu \tuple{\Variable_0,\dots,\Variable_\VarMax} . \tuple{\Formula_0,\dots,\Formula_\VarMax}$
  be a formula of the backward $\mu$-fragment
  with $\BitCount$ propositional constants.
  Without loss of generality,
  we may assume that the subformulas $\Formula_0,\dots,\Formula_\VarMax$
  do not contain any nested modal operators.
  To see this,
  suppose that ${\Formula_i = \bdm \Formula[2]}$.
  Then $\Formula$ is equivalent to
  $\Formula' = \mu \tuple{\Variable_0,\dots,\Variable_i,\dots,\Variable_\VarMax,\Variable[2]}
                 . \tuple{\Formula_0,\dots,\Formula'_i,\dots,\Formula_\VarMax,\Formula[2]}$,
  where $\Variable[2]$ is a fresh propositional variable
  and $\Formula'_i = \bdm \Variable[2]$.
  The operator $\bbx$ and Boolean combinations of $\bdm$ and $\bbx$ are handled analogously.

  We now convert $\Formula$ into an equivalent automaton
  $\Automaton = \tuple{\StateSet,\InitFunc,\TransFunc,\AcceptSet}$
  with state set
  $\StateSet =
   \powerset{\set{\Constant_0,\dots,\Constant_{\BitCount-1},\Variable_0,\dots,\Variable_\VarMax}}$.
  The idea is that
  each node $\Node$ of the input \digraph{} has to remember
  which of the atomic propositions
  $\Constant_0,\dots,\Constant_{\BitCount-1},\Variable_0,\dots,\Variable_\VarMax$
  have, so far, been verified to hold at~$\Node$.
  Therefore,
  we define the initialization function such that
  $\InitFunc(\String) = \setbuilder{\Constant_i}{\String(i)=1}$
  for all $\String \in \Boolean^\BitCount$.
  Let us write
  $\tuple{\State,\NeighborSet} \models \Formula_i$
  to indicate that a pair
  $\tuple{\State,\NeighborSet} \in \StateSet \times \powerset{\StateSet}$
  satisfies a subformula~$\Formula_i$ of~$\Formula$.
  This is the case precisely when $\Formula_i$ holds at any node~$\Node$
  that satisfies exactly the atomic propositions in $\State$
  and whose incoming neighbors satisfy exactly
  the propositions specified by $\NeighborSet$.
  Note that this satisfaction relation is well-defined in our context
  because the nesting depth of modal operators in $\Formula_i$
  is at most $1$.
  With that,
  the transition function of~$\Automaton$ can be succinctly described by
  $\TransFunc(\State,\NeighborSet) =
   \State \cup \setbuilder{\Variable_i}{\tuple{\State,\NeighborSet} \models \Formula_i}$.
  Since $\State \subseteq \TransFunc(\State,\NeighborSet)$,
  we are guaranteed that the automaton is quasi-acyclic.
  Finally,
  the accepting set is given by
  $\AcceptSet = \setbuilder{\State}{\Variable_0 \in \State}$.

  It remains to prove that $\Automaton$ is asynchronous and equivalent to $\Formula$.
  Let
  $\Graph = \tuple{\NodeSet, \EdgeSet, \Labeling}$~be
  an $\BitCount$-bit labeled \digraph{}
  and
  $\vec{\LFixpoint} =
   \tuple{\LFixpoint_0,\dots,\LFixpoint_\VarMax} \in (\powerset{\NodeSet})^{\VarMax+1}$
  be the least fixpoint of the operator~$\Operator$
  associated with $\tuple{\Formula_0,\dots,\Formula_\VarMax}$.
  Due to the asynchrony condition,
  we must consider an arbitrary timing
  $\Timing = \tuple{\Timing_1, \Timing_2, \dots}$
  of~$\Graph$.
  The corresponding run $\Run = \tuple{\Run_0, \Run_1, \dots}$
  of $\Automaton$ on $\Graph$ timed by~$\Timing$
  engenders an infinite sequence
  $\tuple{\vec{\NodeSubset}^0, \vec{\NodeSubset}^1, \dots}$,
  where each tuple
  $\vec{\NodeSubset}^\Time =
   \tuple{\NodeSubset^\Time_0,\dots,\NodeSubset^\Time_\VarMax} \in (\powerset{\NodeSet})^{\VarMax+1}$
  specifies the valuation of every variable $\Variable_i$ at time~$\Time$,
  i.e., $\NodeSubset^\Time_i = \setbuilder{\Node \in \NodeSet}{\Variable_i \in \Run_\Time(\Node)}$.
  Since $\Automaton$ is quasi-acyclic and $\NodeSet$ is finite,
  this sequence must eventually stabilize at some value $\vec{\NodeSubset}^\infty$,
  and each node accepts if and only if it belongs to~$\NodeSubset^\infty_0$.
  Reformulated this way,
  our task is to demonstrate that $\vec{\NodeSubset}^\infty$ equals $\vec{\LFixpoint}$,
  regardless of the timing~$\Timing$.

  “$\vec{\NodeSubset}^\infty \subseteq \vec{\LFixpoint}$”:
  We show by induction that
  $\vec{\NodeSubset}^\Time \subseteq \vec{\LFixpoint}$
  for all $\Time \in \Natural$.
  This obviously holds for $\Time = 0$,
  since $\vec{\NodeSubset}^0 = \tuple{\emptyset,\dots,\emptyset}$.
  Now,
  consider any node $\Node \in \NodeSet$ at an arbitrary time~$\Time$.
  Let $\State$ be the current state of $\Node$
  and $\NeighborSet$ be the set of current states of its incoming neighbors.
  Depending on~$\Timing$,
  it might be the case that $\Node$ actually receives
  some outdated information $\NeighborSet'$
  instead of~$\NeighborSet$.
  However,
  given that the neighbors' previous states
  cannot contain more variables than their current ones
  (by construction),
  and that variables can only occur positively in each~$\Formula_i$,
  we know that
  $\tuple{\State,\NeighborSet'} \models \Formula_i$
  implies
  $\tuple{\State,\NeighborSet} \models \Formula_i$.
  Hence,
  if $\Node$ performs a local transition at time~$\Time$,
  then the only new variables that can be added to its state must lie in
  $\setbuilder{\Variable_i}{\tuple{\State,\NeighborSet} \models \Formula_i}$.
  On a global scale,
  this means that
  $\vec{\NodeSubset}^{\Time+1} \setminus \vec{\NodeSubset}^\Time
   \subseteq \Operator(\vec{\NodeSubset}^\Time)$.
  Furthermore,
  by the induction hypothesis, the monotonicity of $\Operator$,
  and the fact that $\vec{\LFixpoint}$ is a fixpoint,
  we have
  $\Operator(\vec{\NodeSubset}^\Time) \subseteq \Operator(\vec{\LFixpoint})
   = \vec{\LFixpoint}$.
  Putting both together,
  and again relying on the induction hypothesis,
  we obtain~$\vec{\NodeSubset}^{\Time+1} \subseteq \vec{\LFixpoint}$.

  “$\vec{\NodeSubset}^\infty \supseteq \vec{\LFixpoint}$”:
  For the converse direction,
  we make use of the Knaster-Tarski theorem,
  which gives us the equality
  $\vec{\LFixpoint} =
   \bigcap \setbuilder{\vec{\NodeSubset} \in (\powerset{\NodeSet})^{\VarMax+1}}
                      {\Operator(\vec{\NodeSubset}) \subseteq \vec{\NodeSubset}}$.
  With this,
  it suffices to show that
  $\Operator(\vec{\NodeSubset}^\infty) \subseteq \vec{\NodeSubset}^\infty$.
  Consider some time $\Time \in \Natural$
  such that $\vec{\NodeSubset}^{\Time'} = \vec{\NodeSubset}^\infty$
  for all $\Time' \geq \Time$.
  Although we know
  that every node has reached its final state at time $\Time$,
  the FIFO buffers of some edges might still contain obsolete states from previous times.
  However,
  the fairness property of~$\Timing$ guarantees that
  our customized $\popfirst$ operation is executed infinitely often at every edge,
  while the $\pushlast$ operation has no effect
  because all the states remain unchanged.
  Therefore,
  there must be a time $\Time' \geq \Time$
  from which on each buffer contains only the current state
  of its incoming node,
  i.e., $\Run_{\Time''}(\Node[1]\Node[2]) = \Run_{\Time''}(\Node[1])$
  for all $\Time'' \geq \Time'$ and $\Node[1]\Node[2] \in \EdgeSet$.
  Moreover,
  the fairness property of $\Timing$ also ensures that
  every node $\Node$ reevaluates the local transition function~$\TransFunc$ infinitely often,
  based on its own current state $\State$
  and the set $\NeighborSet$ of states in the buffers
  associated with its incoming neighbors.
  As this has no influence on $\Node$'s state,
  we can deduce that
  $\setbuilder{\Variable_i}{\tuple{\State,\NeighborSet} \models \Formula_i} \subseteq \State$.
  Consequently,
  we have
  $\Operator(\vec{\NodeSubset}^{\Time'}) \subseteq \vec{\NodeSubset}^{\Time'}$,
  which is equivalent to
  $\Operator(\vec{\NodeSubset}^\infty) \subseteq \vec{\NodeSubset}^\infty$.
\end{proof}
\section{Capturing asynchronous runs using least fixpoints}
\label{sec:automata-to-logic}
This section is dedicated to proving
the converse direction of the main result,
which will allow us to translate any
quasi-acyclic lossless-asynchronous automaton
into an equivalent formula of the backward $\mu$-fragment
(see Proposition~\ref{prp:automata-to-logic}).
Our proof builds on two concepts:
the invariance of distributed automata under backward bisimulation
(stated in Proposition~\ref{prp:bisimulation-invariance})
and an ad-hoc relation “$\enables$”
that captures the possible behaviors
of a fixed lossless-asynchronous automaton~$\Automaton$
(in a specific sense described in Lemma~\ref{lem:enables-relation}).

We start with the notion of backward bisimulation,
which is defined like the standard notion of bisimulation
(see, e.g., \cite[Def.~2.16]{BlackburnRV02} or \cite[Def.~5]{BlackburnB07}),
except that edges are followed in the backward direction.
Formally,
a \defd{backward bisimulation}
between two $\BitCount$-bit labeled \digraph{s}
$\Graph = \tuple{\NodeSet, \EdgeSet, \Labeling}$
and
$\Graph' = \tuple{\NodeSet', \EdgeSet', \Labeling'}$
is a binary relation
$\Bisimulation \subseteq \NodeSet \times \NodeSet'$
that fulfills the following conditions
for all $\Node[2]\Node[2]' \in \Bisimulation$:
\begin{enumerate}
\item $\Labeling(\Node[2]) = \Labeling'(\Node[2]')$,
\item if $\Node[1]\Node[2] \in \EdgeSet$,
  then there exists $\Node[1]' \in \NodeSet'$
  such that
  $\Node[1]'\Node[2]' \in \EdgeSet'$
  and
  $\Node[1]\Node[1]' \in \Bisimulation$,\,
  and, conversely,
\item if $\Node[1]'\Node[2]' \in \EdgeSet'$,
  then there exists $\Node[1] \in \NodeSet$
  such that
  $\Node[1]\Node[2] \in \EdgeSet$
  and
  $\Node[1]\Node[1]' \in \Bisimulation$.
\end{enumerate}
We say that
the pointed \digraph{s}
$\tuple{\Graph,\Node}$ and $\tuple{\Graph',\Node'}$
are \defd{backward bisimilar}
if there exists such a backward bisimulation $\Bisimulation$
relating $\Node$ and $\Node'$.
It is easy to see that
distributed automata cannot distinguish
between backward bisimilar structures:

\begin{proposition}
  \label{prp:bisimulation-invariance}
  Distributed automata are invariant under backward bisimulation.
  That is,
  for every automaton $\Automaton$,
  if two pointed \digraph{s}
  $\tuple{\Graph,\Node[2]}$ and $\tuple{\Graph',\Node[2]'}$
  are backward bisimilar,
  then
  $\Automaton$ accepts $\tuple{\Graph,\Node[2]}$
  if and only if
  it accepts $\tuple{\Graph',\Node[2]'}$.
\end{proposition}
\begin{proof}
  Let $\Bisimulation$ be a backward bisimulation
  between $\Graph$ and $\Graph'$
  such that $\Node[2]\Node[2]' \in \Bisimulation$.
  Since acceptance is defined with respect to
  the synchronous behavior of the automaton,
  we need only consider the synchronous runs
  $\Run = \tuple{\Run_0, \Run_1, \dots}$ and
  $\Run' = \tuple{\Run'_0, \Run'_1, \dots}$
  of~$\Automaton$ on~$\Graph$ and~$\Graph'$, respectively.
  Now,
  given that the FIFO buffers on the edges of the \digraph{s}
  merely contain the current state of their incoming node,
  it is straightforward to prove by induction on $\Time$ that
  every pair of nodes $\Node[1]\Node[1]' \in \Bisimulation$ satisfies
  $\Run_\Time(\Node[1]) = \Run'_\Time(\Node[1]')$
  for all $\Time \in \Natural$.
\end{proof}

We now turn to the mentioned relation “$\enables$”,
which is defined with respect to a fixed automaton.
For the remainder of this section,
let~$\Automaton$ denote an automaton
$\tuple{\StateSet,\InitFunc,\TransFunc,\AcceptSet}$,
and let~$\TraceSet$ denote its set of traces.
The relation
${\enables} \subseteq (\powerset{\TraceSet} \times \TraceSet)$
specifies whether,
in a lossless-asynchronous environment,
a given trace~$\Trace$ can be traversed by a node
whose incoming neighbors traverse the traces of a given set~$\NeighborHistory$.
Loosely speaking,
the intended meaning of
\defd{$\NeighborHistory \enables \Trace$} (“$\NeighborHistory$~enables~$\Trace$”)
is the following:
Take an appropriately chosen \digraph{}
under some lossless-asynchronous timing~$\Timing$,
and observe the corresponding run of $\Automaton$
up to a specific time~$\Time$;
if node~$\Node$ was initially in state $\Trace.\first$
and at time~$\Time$ it has \emph{seen} its incoming neighbors
traversing precisely the traces in~$\NeighborHistory$,
then it is possible for~$\Timing$ to be such that
at time~$\Time$,
node $\Node$ has traversed exactly the trace~$\Trace$.
This relation can be defined inductively:
As the base case,   
we specify that for every
$\State \in \StateSet$ and $\NeighborSet \subseteq \StateSet$,
we have
${\NeighborSet \enables \State.\pushlast(\TransFunc(\State,\NeighborSet))}$.
For the inductive clause,
consider a trace $\Trace[2] \in \TraceSet$
and two finite (possibly equal) sets of traces
$\NeighborHistory, \NeighborHistory' \subseteq \TraceSet$
such that the traces in $\NeighborHistory'$ can be obtained
by appending at most one state to the traces in $\NeighborHistory$.
More precisely,
if $\Trace[1] \in \NeighborHistory$,
then $\Trace[1].\pushlast(\State[1]) \in \NeighborHistory'$
for some $\State[1] \in \StateSet$,
and conversely,
if $\Trace[1]' \in \NeighborHistory'$,
then $\Trace[1]' = \Trace[1].\pushlast(\Trace[1]'.\last)$
for some $\Trace[1] \in \NeighborHistory$.
We shall denote this auxiliary relation by
\defd{$\NeighborHistory \becomes \NeighborHistory'$}.
If it holds,
then
$\NeighborHistory \enables \Trace[2]$
implies
$\NeighborHistory' \enables \Trace[2].\pushlast(\State[2])$,
where
$\State[2] =
 \TransFunc(\Trace[2].\last,
            \setbuilder{\Trace[1]'.\last}{\Trace[1]' \in \NeighborHistory'})$.

The next step is to show (in Lemma~\ref{lem:enables-relation})
that our definition of “$\enables$”
does indeed capture the intuition given above.
To formalize this,
we first introduce two further pieces of terminology.

First,
the notions of configuration and run can be enriched
to facilitate discussions about the past.
Let $\Run = \tuple{\Run_0, \Run_1, \dots}$
be a run of $\Automaton$ on a \digraph{}
$\Graph = \tuple{\NodeSet, \EdgeSet, \Labeling}$
(timed by some timing~$\Timing$).
The corresponding \defd{enriched run} is the sequence
$\RichRun = \tuple{\RichRun_0, \RichRun_1, \dots}$
of \defd{enriched configurations}
that we obtain from $\Run$
by requiring each node to remember the entire trace it has traversed so far.
Formally,
for $\Time \in \Natural$, $\Node \in \NodeSet$ and $\Edge \in \EdgeSet$,
\begin{equation*}
  \RichRun_0(\Node) = \Run_0(\Node),
  \qquad
  \RichRun_{\Time+1}(\Node) = \RichRun_\Time(\Node).\pushlast(\Run_{\Time+1}(\Node))
  \qquad \text{and} \qquad
  \RichRun_\Time(\Edge) = \Run_\Time(\Edge).
\end{equation*}

Second,
we will need to consider finite segments of timings and enriched runs.
A \defd{lossless-asynchronous timing segment} of a \digraph{}~$\Graph$
is a finite sequence
$\Timing = \tuple{\Timing_1, \dots, \Timing_\EndTime}$
that could be extended to a whole lossless-asynchronous timing
$\tuple{\Timing_1, \dots, \Timing_\EndTime, \Timing_{\EndTime+1}, \dots}$.
Likewise,
for an initial enriched configuration~$\RichRun_0$ of~$\Graph$,
the corresponding \defd{enriched run segment} timed by $\Timing$
is the sequence $\tuple{\RichRun_0, \dots, \RichRun_\EndTime}$,
where each $\RichRun_{\Time+1}$ is computed
from $\RichRun_\Time$ and $\Timing_{\Time+1}$
in the same way as for an entire enriched run.

Equipped with the necessary terminology,
we can now state and prove a (slightly technical) lemma
that will allow us to derive benefit from the relation “$\enables$”.
This lemma essentially states that
if $\NeighborHistory \enables \Trace[2]$ holds
and we are given enough nodes
that traverse the traces in~$\NeighborHistory$,
then we can take those nodes as the incoming neighbors
of a new node~$\Node[2]$
and delay the messages received by~$\Node[2]$
in such a way that $\Node[2]$ traverses $\Trace[2]$,
without losing any messages.

\begin{lemma}
  \label{lem:enables-relation}
  For every trace $\Trace[2] \in \TraceSet$
  and every finite (possibly empty) set of traces
  $\NeighborHistory = \set{\Trace[1]_1, \dots, \Trace[1]_\ell} \subseteq \TraceSet$
  that satisfy the relation $\NeighborHistory \enables \Trace[2]$,
  there exist lower bounds $m_1, \dots, m_\ell \in \Positive$
  such that the following statement holds true:

  For any $n_1, \dots, n_\ell \in \Positive$ satisfying $n_i \geq m_i$,
  let $\Graph$ be a \digraph{} consisting of
  the nodes $\tuple{\Node[1]_i^j}_{i,j}$ and $\Node[2]$,
  and the edges $\tuple{\Node[1]_i^j\Node[2]}_{i,j}$,
  with index ranges $1 \leq i \leq \ell$ and $1 \leq j \leq n_i$.
  If we start from the enriched configuration~$\RichRun_0$ of~$\Graph$,
  where
  \begin{equation*}
    \RichRun_0(\Node[1]_i^j) = \Trace[1]_i,
    \qquad
    \RichRun_0(\Node[1]_i^j\Node[2]) = \Trace[1]_i
    \qquad \text{and} \qquad
    \RichRun_0(\Node[2]) = \Trace[2].\first,
  \end{equation*}
  then we can construct a (nonempty) lossless-asynchronous timing segment
  $\Timing = \tuple{\Timing_1, \dots, \Timing_\EndTime}$ of~$\Graph$,
  where
  $\Timing_\Time(\Node[1]_i^j) = 0$ and
  $\Timing_\Time(\Node[2]) = 1$
  for $1 \leq \Time \leq \EndTime$,
  such that the corresponding enriched run segment
  $\RichRun = \tuple{\RichRun_0, \dots, \RichRun_\EndTime}$ timed by $\Timing$
  satisfies
  \begin{equation*}
    \RichRun_{\EndTime-1}(\Node[1]_i^j\Node[2]) = \Trace[1]_i.\last
    \qquad \text{and} \qquad
    \RichRun_\EndTime(\Node[2]) = \Trace[2].
  \end{equation*}
\end{lemma}
\begin{proof}
  We proceed by induction on the definition of~“$\enables$”.
  In the base case,
  where
  $\NeighborHistory = \set{\State[1]_1, \dots, \State[1]_\ell} \subseteq \StateSet$
  and
  $\Trace[2] = \State[2].\pushlast(\TransFunc(\State[2],\NeighborHistory))$
  for some $\State[2] \in \StateSet$,
  the statement holds with $m_1 = \dots = m_\ell = 1$.
  This is witnessed by a timing segment $\Timing = \tuple{\Timing_1}$,
  where
  $\Timing_1(\Node[1]_i^j) = 0$,\,
  $\Timing_1(\Node[2]) = 1$,
  and $\Timing_1(\Node[1]_i^j\Node[2])$ can be chosen as desired.

  For the inductive step,
  we assume that the statement holds for
  $\Trace[2]$ and
  $\NeighborHistory = \set{\Trace[1]_1, \dots, \Trace[1]_\ell}$
  with some values $m_1, \dots, m_\ell$.
  Now consider any other set of traces
  $\NeighborHistory' = \set{\Trace[1]'_1, \dots, \Trace[1]'_{\ell'}}$
  such that $\NeighborHistory \becomes \NeighborHistory'$,
  and let $\Trace[2]' = \Trace[2].\pushlast(\State[2])$,
  where
  $\State[2] =
   \TransFunc(\Trace[2].\last,
              \setbuilder{\Trace[1]'_k.\last}{\Trace[1]'_k \in \NeighborHistory'})$.
  Since $\NeighborHistory \enables \Trace[2]$,
  we have $\NeighborHistory' \enables \Trace[2]'$.
  The remainder of the proof consists in showing that
  the statement also holds for $\Trace[2]'$ and $\NeighborHistory'$
  with some large enough integers $m'_1, \dots, m'_{\ell'}$.
  Let us fix
  $m'_k = \sum \setbuilder{m_i}{\Trace[1]_i.\pushlast(\Trace[1]'_k.\last) = \Trace[1]'_k}$.
  (As there is no need to find minimal values, we opt for easy expressibility.)

  Given any numbers $n'_1, \dots, n'_{\ell'}$ with $n'_k \geq m'_k$,
  we choose suitable values $n_1, \dots, n_\ell$ with $n_i \geq m_i$,
  and consider the corresponding \digraph{} $\Graph$ described in the lemma.
  Because we have $\NeighborHistory \becomes \NeighborHistory'$,
  we can assign to each node $\Node[1]_i^j$ a state $\State[1]_i^j$
  such that
  $\Trace[1]_i.\pushlast(\State[1]_i^j) \in \NeighborHistory'$.
  Moreover,
  provided our choice of $n_1, \dots, n_\ell$ was adequate,
  we can also ensure that for each $\Trace[1]'_k \in \NeighborHistory'$,
  there are exactly $n'_k$ nodes $\Node[1]_i^j$ such that
  $\Trace[1]_i.\pushlast(\State[1]_i^j) = \Trace[1]'_k$.
  (Note that nodes with distinct traces
  $\Trace[1]_i, \Trace[1]_{i'} \in \NeighborHistory$
  might be mapped to the same trace
  $\Trace[1]'_k \in \NeighborHistory'$,
  in case $\Trace[1]_{i'} = \Trace[1]_i\State[1]_i^j$.)
  It is straightforward to verify
  that such a choice of numbers and such an assignment of states
  are always possible,
  given the lower bounds $m'_1, \dots, m'_{\ell'}$ specified above.

  Let us now consider
  the lossless-asynchronous timing segment
  $\Timing = \tuple{\Timing_1, \dots, \Timing_\EndTime}$
  and the corresponding enriched run segment
  $\RichRun = \tuple{\RichRun_0, \dots, \RichRun_\EndTime}$
  provided by the induction hypothesis.
  Since the $\popfirst$ operation has no effect on a trace of length $1$,
  we may assume without loss of generality
  that
  $\Timing_\Time(\Node[1]_i^j\Node[2]) = 0$
  if
  $\RichRun_{\Time-1}(\Node[1]_i^j\Node[2])$ has length $1$,
  for $\Time < \EndTime$.
  Consequently,
  if we start from the alternative enriched configuration~$\RichRun'_0$,
  where
  \begin{equation*}
    \RichRun'_0(\Node[1]_i^j) = \Trace[1]_i.\pushlast(\State[1]_i^j),
    \qquad
    \RichRun'_0(\Node[1]_i^j\Node[2]) = \Trace[1]_i.\pushlast(\State[1]_i^j)
    \qquad \text{and} \qquad
    \RichRun'_0(\Node[2]) = \Trace[2].\first,
  \end{equation*}
  then the corresponding enriched run segment
  $\tuple{\RichRun'_0, \dots, \RichRun'_\EndTime}$ timed by $\Timing$
  can be derived from~$\RichRun$ by simply applying “$\pushlast(\State[1]_i^j)$”
  to
  $\RichRun_\Time(\Node[1]_i^j)$ and $\RichRun_\Time(\Node[1]_i^j\Node[2])$,
  for $\Time < \EndTime$.
  We thus get
  \begin{equation*}
    \RichRun'_{\EndTime-1}(\Node[1]_i^j\Node[2]) = \Trace[1]_i.\last.\pushlast(\State[1]_i^j)
    \qquad \text{and} \qquad
    \RichRun'_\EndTime(\Node[2]) = \Trace[2].
  \end{equation*}
  We may also assume without loss of generality
  that $\Timing_\EndTime(\Node[1]_i^j\Node[2]) = 1$
  if $\RichRun'_{\EndTime-1}(\Node[1]_i^j\Node[2])$ has length~$2$,
  since this does not affect $\RichRun$
  and lossless-asynchrony is ensured by $\Timing_\EndTime(\Node[2]) = 1$.
  Hence,
  it suffices to extend $\Timing$ by an additional map $\Timing_{\EndTime+1}$,
  where
  $\Timing_{\EndTime+1}(\Node[1]_i^j) = 0$,\,
  $\Timing_{\EndTime+1}(\Node[2]) = 1$,
  and $\Timing_{\EndTime+1}(\Node[1]_i^j\Node[2])$ can be chosen as desired.
  The resulting enriched run segment
  $\tuple{\RichRun'_0, \dots, \RichRun'_{\EndTime+1}}$
  satisfies
  \begin{equation*}
    \RichRun'_\EndTime(\Node[1]_i^j\Node[2]) = \State[1]_i^j = \Trace[1]'_k.\last
    \;\; \text{(for some $\Trace[1]'_k \in \NeighborHistory'$)}
    \qquad \text{and} \qquad
    \RichRun'_{\EndTime+1}(\Node[2]) = \Trace[2].\pushlast(\State[2]) = \Trace[2]'.
    \qedhere
  \end{equation*}
\end{proof}

Finally,
we can put the pieces together
and prove the converse direction of Theorem~\ref{thm:main-result}:

\begin{proposition}[$\QLA \subseteq \SigmaOne$]
  \label{prp:automata-to-logic}
  For every quasi-acyclic lossless-asynchronous automaton,
  we can effectively construct
  an equivalent formula of the backward $\mu$-fragment.
\end{proposition}
\begin{proof}
  Assume that
  $\Automaton = \tuple{\StateSet,\InitFunc,\TransFunc,\AcceptSet}$
  is a quasi-acyclic lossless-asynchronous automaton
  with $\BitCount$-bit input.
  Since it is quasi-acyclic,
  its set of traces $\TraceSet$ is finite,
  and thus we can afford to introduce
  a separate propositional variable $\Variable_{\Trace}$
  for each trace $\Trace \in \TraceSet$.
  Making use of the relation~“$\enables$”,
  we convert~$\Automaton$ into an equivalent formula
  $\Formula =
   \mu \bigl[ \Variable_0, \tuple{\Variable_{\Trace}}_{\Trace \in \TraceSet} \bigr] .
       \bigl[ \Formula_0, \tuple{\Formula_{\Trace}}_{\Trace \in \TraceSet} \bigr]$
  of the backward $\mu$-fragment,
  where
  \begin{align}
    \Formula_0 &=
    \smashoperator[r]{\bigvee_{\substack{\Trace \in \TraceSet \\ \Trace\ldotp\last \in \AcceptSet}}}
    \Variable_{\Trace},
    \tag{a} \label{eq:accept} \\
    \Formula_{\State} &=
    \smashoperator[r]{\bigvee_{\substack{\String \in \Boolean^\BitCount \\ \InitFunc(\String)=\State}}} \:\:
    \Bigl(
      \smashoperator[r]{\bigwedge_{\String(i)=1}} \Constant_i
      \,\land\!     
      \smashoperator[r]{\bigwedge_{\String(i)=0}} \lnot \Constant_i
    \Bigr)
    \qquad \text{for each $\State \in \StateSet$,\quad and}
    \tag{b} \label{eq:init} \\
    \Formula_{\Trace[2]} &=
    \Variable_{\Trace[2]\ldotp\first}
    \,\land
    \bigvee_{\substack{\NeighborHistory \subseteq \TraceSet \\ \NeighborHistory \:\!\enables\:\! \Trace[2]}}
    \Bigl(
      \bigl(\, \smashoperator{\bigwedge_{\Trace[1] \in \NeighborHistory}} \bdm \Variable_{\Trace[1]} \bigr)
      \land
      \bigl( \bbx \smashoperator{\bigvee_{\Trace[1] \in \NeighborHistory}} \Variable_{\Trace[1]} \bigr)
    \Bigr)
    \qquad \text{for each $\Trace[2] \in \TraceSet$ with $\length{\Trace[2]} \geq 2$.}
    \tag{c} \label{eq:transition}
  \end{align}
  Note that this formula can be constructed effectively
  because an inductive computation of~“$\enables$” must terminate
  after at most $\card{\TraceSet} \cdot 2^{\card{\TraceSet}}$ iterations.

  To prove that $\Formula$ is indeed equivalent to $\Automaton$,
  let us consider an arbitrary $\BitCount$-bit labeled \digraph{}
  $\Graph = \tuple{\NodeSet, \EdgeSet, \Labeling}$
  and the corresponding least fixpoint
  $\vec{\LFixpoint} =
   \tuple{\LFixpoint_0, \tuple{\LFixpoint_{\Trace}}_{\Trace \in \TraceSet}}
   \in (\powerset{\NodeSet})^{\card{\TraceSet}+1}$
  of the operator $\Operator$ associated with
  $\tuple{\Formula_0, \tuple{\Formula_{\Trace}}_{\Trace \in \TraceSet}}$.

  The easy direction is to show that
  for all nodes $\Node \in \NodeSet$,
  if $\Automaton$ accepts $\tuple{\Graph,\Node}$,
  then $\tuple{\Graph,\Node}$ satisfies $\Formula$.
  For that,
  it suffices to consider the synchronous enriched run
  $\RichRun = \tuple{\RichRun_0, \RichRun_1, \dots}$
  of~$\Automaton$ on~$\Graph$.
  (Any other run timed by a lossless-asynchronous timing
  would exhibit the same acceptance behavior.)
  As in the proof of Proposition~\ref{prp:bisimulation-invariance},
  we can simply ignore the FIFO buffers on the edges of~$\Graph$
  because $\RichRun_\Time(\Node[1]\Node[2]) = \RichRun_\Time(\Node[1]).\last$.
  Using this,
  a straightforward induction on $\Time$ shows that
  every node $\Node[2] \in \NodeSet$ satisfies
  ${\setbuilder{\RichRun_\Time(\Node[1])}{\Node[1]\Node[2] \in \EdgeSet}
    \enables \RichRun_{\Time+1}(\Node[2])}$
  for all $\Time \in \Natural$.
  (For~$\Time = 0$,
   the claim follows from the base case of the definition of~“$\enables$”;
   for the step from~$\Time$ to~$\Time+1$,
   we can immediately apply the inductive clause of the definition.)
  This in turn allows us to prove that
  each node~$\Node[2]$ is contained in all the components of $\vec{\LFixpoint}$
  that correspond to a trace traversed by~$\Node[2]$ in~$\RichRun$,
  i.e., $\Node[2] \in \LFixpoint_{\RichRun_\Time(\Node[2])}$ for all $\Time \in \Natural$.
  Naturally, we proceed again by induction:
  For $\Time = 0$,
  we have $\RichRun_0(\Node[2]) = \InitFunc(\Labeling(\Node[2])) \in \StateSet$,
  hence the subformula $\Formula_{\RichRun_0(\Node[2])}$
  defined in equation~\eqref{eq:init}
  holds at~$\Node[2]$,
  and thus $\Node[2] \in \LFixpoint_{\RichRun_0(\Node[2])}$.
  For the step from~$\Time$ to~$\Time+1$,
  we need to distinguish two cases.
  If $\RichRun_{\Time+1}(\Node[2])$ is of length $1$,
  then it is equal to $\RichRun_\Time(\Node[2])$,
  and there is nothing new to prove.
  Otherwise,
  we must consider the appropriate subformula
  $\Formula_{\RichRun_{\Time+1}(\Node[2])}$
  given by equation~\eqref{eq:transition}.
  We already know from the base case that the conjunct
  $\Variable_{\RichRun_{\Time+1}(\Node[2])\ldotp\first} = \Variable_{\RichRun_0(\Node[2])}$
  holds at $\Node[2]$,
  with respect to any variable assignment
  that interprets each $\Variable_{\Trace}$ as $\LFixpoint_{\Trace}$.
  Furthermore,
  by the induction hypothesis,
  $\Variable_{\RichRun_\Time(\Node[1])}$
  holds at every incoming neighbor $\Node[1]$ of $\Node[2]$.
  Since
  ${\setbuilder{\RichRun_\Time(\Node[1])}{\Node[1]\Node[2] \in \EdgeSet}
    \enables \RichRun_{\Time+1}(\Node[2])}$,
  we conclude that the second conjunct of $\Formula_{\RichRun_{\Time+1}(\Node[2])}$
  must also hold at $\Node[2]$,
  and thus $\Node[2] \in \LFixpoint_{\RichRun_{\Time+1}(\Node[2])}$.
  Finally,
  assuming $\Automaton$ accepts $\tuple{\Graph,\Node}$,
  we know by definition that
  $\RichRun_\Time(\Node).\last \in \AcceptSet$ for some $\Time \in \Natural$.
  Since $\Node \in \LFixpoint_{\RichRun_\Time(\Node)}$,
  this implies that
  the subformula~$\Formula_0$ defined in equation~\eqref{eq:accept}
  holds at $\Node$,
  and therefore that $\tuple{\Graph,\Node}$ satisfies $\Formula$.

  For the converse direction of the equivalence,
  we have to overcome the difficulty
  that~$\Formula$ is more permissive than~$\Automaton$,
  in the sense that
  a node $\Node$ might lie in $\LFixpoint_{\Trace}$,
  and yet not be able to follow the trace $\Trace$ under any timing of $\Graph$.
  Intuitively,
  the reason why we still obtain an equivalence
  is that $\Automaton$ cannot take advantage
  of all the information provided by any particular run,
  because it must ensure that for \emph{all} \digraph{s},
  its acceptance behavior is independent of the timing.
  It turns out that
  even if $\Node$ cannot traverse $\Trace$,
  some other node~$\Node'$ in an indistinguishable \digraph{}
  will be able to do so.
  More precisely, we will show that
  \begin{equation}
    \parbox{0.89\textwidth}{
      if $\Node \in \LFixpoint_{\Trace}$,
      then there exists a pointed \digraph{} $\tuple{\Graph',\Node'}$,
      backward bisimilar to $\tuple{\Graph,\Node}$,
      and a lossless-asynchronous timing $\Timing'$ of $\Graph'$,
      such that $\RichRun'_\Time(\Node') = \Trace$
      for some $\Time \in \Natural$,
    }
    \tag{$\ast$} \label{stm:traversable}
  \end{equation}
  where $\RichRun'$ is the enriched run
  of $\Automaton$ on $\Graph'$ timed by $\Timing'$.
  Now suppose that $\tuple{\Graph,\Node}$ satisfies $\Formula$.
  By equation~\eqref{eq:accept},
  this means that $\Node \in \LFixpoint_{\Trace}$
  for some trace $\Trace$ such that $\Trace.\last \in \AcceptSet$.
  Consequently,
  $\Automaton$~accepts
  the pointed \digraph{} $\tuple{\Graph',\Node'}$
  postulated in~\eqref{stm:traversable},
  based on the claim that~$\Node'$ traverses~$\Trace$ under timing~$\Timing'$
  and the fact that $\Automaton$ is lossless-asynchronous.
  Since $\tuple{\Graph,\Node}$ and $\tuple{\Graph',\Node'}$
  are backward bisimilar,
  it follows from Proposition~\ref{prp:bisimulation-invariance}
  that~$\Automaton$ also accepts~$\tuple{\Graph,\Node}$.

  It remains to verify \eqref{stm:traversable}.
  We achieve this by
  computing the least fixpoint~$\vec{\LFixpoint}$ inductively
  and proving the statement by induction on the sequence of approximants
  $\tuple{\vec{\LFixpoint}^0, \vec{\LFixpoint}^1, \dots}$.
  Note that we do not need to consider the limit case,
  since $\vec{\LFixpoint} = \vec{\LFixpoint}^n$ for some $n \in \Natural$.

  The base case is trivially true
  because all the components of $\vec{\LFixpoint}^0$ are empty.
  Furthermore,
  if~$\Trace[2]$~consists of a single state~$\State$,
  then we do not even need to argue by induction,
  as it is evident from equation~\eqref{eq:init} that
  for all $j \geq 1$,
  node $\Node[2]$ lies in $\LFixpoint^j_{\State}$
  precisely when
  $\InitFunc(\Labeling(\Node[2]))= \State$.
  It thus suffices to set $\tuple{\Graph',\Node[2]'} = \tuple{\Graph,\Node[2]}$
  and choose the timing~$\Timing'$ arbitrarily.
  Clearly,
  we have $\RichRun'_0(\Node[2]') = \InitFunc(\Labeling(\Node[2]))= \State$
  if $\Node[2] \in \LFixpoint^j_{\State}$.

  On the other hand,
  if~$\Trace[2]$ is of length at least $2$,
  we must assume that
  statement~\eqref{stm:traversable} holds for the components of~$\vec{\LFixpoint}^j$
  in order to prove it for $\LFixpoint^{j+1}_{\Trace[2]}$.
  To this end,
  consider an arbitrary node $\Node[2] \in \LFixpoint^{j+1}_{\Trace[2]}$.
  By the first conjunct in~\eqref{eq:transition}
  and the preceding remarks regarding the trivial cases,
  we know that $\InitFunc(\Labeling(\Node[2])) = \Trace[2].\first$
  (and incidentally that $j \geq 1$).
  Moreover,
  the second conjunct ensures
  the existence of a (possibly empty) set of traces $\NeighborHistory$
  that satisfies
  $\NeighborHistory \enables \Trace[2]$
  and that represents
  a “projection” of~$\Node[2]$'s incoming neighborhood at stage~$j$.
  By the latter we mean that
  for all $\Trace[1] \in \NeighborHistory$,
  there exists $\Node[1] \in \NodeSet$
  such that
  $\Node[1]\Node[2] \in \EdgeSet$ and $\Node[1] \in \LFixpoint^j_{\Trace[1]}$,
  and conversely,
  for all $\Node[1] \in \NodeSet$ with $\Node[1]\Node[2] \in \EdgeSet$,
  there exists $\Trace[1] \in \NeighborHistory$
  such that $\Node[1] \in \LFixpoint^j_{\Trace[1]}$.

  Now,
  for each trace $\Trace[1] \in \NeighborHistory$
  and each incoming neighbor~$\Node[1]$ of~$\Node[2]$
  that is contained in $\LFixpoint^j_{\Trace[1]}$,
  the induction hypothesis provides us with a pointed digraph
  $\tuple{\Graph'_{\Node[1]:\Trace[1]}, \Node[1]'_{\Trace[1]}}$
  and a corresponding timing $\Timing'_{\Node[1]:\Trace[1]}$,
  as described in~\eqref{stm:traversable}.
  We make $n_{\Node[1]:\Trace[1]} \in \Natural$ distinct copies
  of each such \digraph{}~$\Graph'_{\Node[1]:\Trace[1]}$.
  From this,
  we construct $\Graph' = \tuple{\NodeSet', \EdgeSet', \Labeling'}$
  by taking the disjoint union of all the $\sum n_{\Node[1]:\Trace[1]}$ \digraph{s},
  and adding a single new node $\Node[2]'$
  with $\Labeling'(\Node[2]') = \Labeling(\Node[2])$,
  together with all the edges of the form~$\Node[1]'_{\Trace[1]}\Node[2]'$
  (i.e., one such edge for each copy of every~$\Node[1]'_{\Trace[1]}$).
  Given that every $\tuple{\Graph'_{\Node[1]:\Trace[1]}, \Node[1]'_{\Trace[1]}}$
  is backward bisimilar to $\tuple{\Graph,\Node[1]}$,
  we can guarantee that the same holds
  for $\tuple{\Graph',\Node[2]'}$ and $\tuple{\Graph,\Node[2]}$
  by choosing the numbers of \digraph{} copies in $\Graph'$ such that
  each incoming neighbor~$\Node[1]$ of~$\Node[2]$ is represented
  by at least one incoming neighbor of~$\Node[2]'$.
  That is, for every~$\Node[1]$,
  we require that ${n_{\Node[1]:\Trace[1]} \geq 1}$ for some~$\Trace[1]$.

  Finally,
  we construct a suitable lossless-asynchronous timing $\Timing'$ of~$\Graph'$,
  which proceeds in two phases
  to make $\Node[2]'$ traverse $\Trace[2]$
  in the corresponding enriched run $\RichRun'$.
  In the first phase,
  where $0 < \Time \leq \Time_1$,
  node~$\Node[2]'$ remains inactive,
  which means that
  every~$\Timing_\Time$ assigns~$0$ to~$\Node[2]'$ and its incoming edges.
  The state of~$\Node[2]'$ at time~$\Time_1$ is thus still $\Trace[2].\first$.
  Meanwhile,
  in every copy of each \digraph{}~$\Graph'_{\Node[1]:\Trace[1]}$,
  the nodes and edges behave according to timing~$\Timing'_{\Node[1]:\Trace[1]}$
  until the respective copy of~$\Node[1]'_{\Trace[1]}$
  has completely traversed~$\Trace[1]$,
  whereupon the entire subgraph becomes inactive.
  By choosing $\Time_1$ large enough,
  we make sure that the FIFO buffer on each edge
  of the form~$\Node[1]'_{\Trace[1]}\Node[2]'$
  contains precisely $\Trace[1]$ at time~$\Time_1$.
  In the second phase,
  which lasts from~$\Time_1 + 1$ to~$\Time_2$,
  the only active parts of $\Graph'$ are $\Node[2]'$ and its incoming edges.
  Since the number~$n_{\Node[1]:\Trace[1]}$ of copies
  of each \digraph{}~$\Graph'_{\Node[1]:\Trace[1]}$
  can be chosen as large as required,
  we stipulate that for every trace~$\Trace[1] \in \NeighborHistory$,
  the sum of $n_{\Node[1]:\Trace[1]}$ over all $\Node[1]$
  exceeds the lower bound~$m_{\Trace[1]}$ that is associated with $\Trace[1]$
  when invoking Lemma~\ref{lem:enables-relation}
  for~$\Trace[2]$ and~$\NeighborHistory$.
  Applying that lemma,
  we obtain a lossless-asynchronous timing segment
  of the subgraph induced by $\Node[2]'$ and its incoming neighbors.
  This segment determines our timing~$\Timing'$ between~$\Time_1 + 1$ and~$\Time_2$
  (the other parts of~$\Graph'$ being inactive),
  and gives us $\RichRun'_{\Time_2}(\Node[2]') = \Trace[2]$, as desired.
  Naturally,
  the remainder of~$\Timing'$, starting at~$\Time_2 + 1$, can be chosen arbitrarily,
  so long as it satisfies the properties of a lossless-asynchronous timing.

  As a closing remark,
  note that the pointed \digraph{}~$\tuple{\Graph',\Node[2]'}$
  constructed above is very similar to the standard unraveling
  of $\tuple{\Graph,\Node[2]}$ into a (possibly infinite) tree.
  (The set of nodes of that tree-unraveling is precisely
  the set of all directed paths in $\Graph$ that start at $\Node[2]$;
  see, e.g., \cite[Def.~4.51]{BlackburnRV02} or \cite[\S~3.2]{BlackburnB07}).
  However, there are a few differences:
  First,
  we do the unraveling backwards,
  because we want to generate a backward bisimilar structure,
  where all the edges point toward the root.
  Second,
  we may duplicate the incoming neighbors (i.e., children) of each node in the tree,
  in order to satisfy the lower bounds imposed by Lemma~\ref{lem:enables-relation}.
  Third,
  we stop the unraveling process at a finite depth
  (not necessarily the same for each subtree),
  and place a copy of the original \digraph{}~$\Graph$ at every~leaf.
\end{proof}

\paragraph*{Acknowledgments}

I would like to thank Olivier Carton,
my PhD supervisor,
for many pleasant discussions and constructive comments.
This work is supported by the
\href{http://delta.labri.fr/}{DeLTA project} (ANR-16-CE40-0007).

\bibliography{mu-paper.bib}

\end{document}